\documentstyle{AGILEpsr} 
 
\input epsf.sty 
\input epsfig.sty 
\input psfig.sty 
 
\def \AAP #1 #2 {{\em Astron. Astrophys.\/} {\bf #1}, #2} 
\def \AAL #1 #2 {{\em Astron. Astrophys. Lett.\/} {\bf #1}, L#2} 
\def \AAR #1 #2 {{\em Astron. Astrophys. Rev.\/} {\bf #1}, #2} 
\def \AAS #1 #2 {{\em Astron. Astrophys. Suppl. Ser.\/} {\bf #1}, #2} 
\def \AJ #1 #2 {{\em Astron. J.\/} {\bf #1}, #2} 
\def \ANNREV #1 #2 {{\em Ann. Rev. Astron. Astrophys.\/} {\bf #1}, #2} 
\def \APJ #1 #2 {{\em Astrophys. J.\/} {\bf #1}, #2} 
\def \APJL #1 #2 {{\em Astrophys. J. Lett.\/} {\bf #1}, L#2} 
\def \APJS #1 #2 {{\em Astrophys. J. Suppl.\/} {\bf #1}, #2} 
\def \APSS #1 #2 {{\em Astrophys. Space Sci.\/} {\bf #1}, #2} 
\def \ASR #1 #2 {{\em Adv. Space Res.\/} {\bf #1}, #2} 
\def \BAIC #1 #2 {{\em Bull. Astron. Inst. Czechosl.\/} {\bf #1}, #2} 
\def \JSQRT #1 #2 {{\em J. Quant. Spectrosc. Radiat. Transfer\/} {\bf #1}, #2} 
\def \MN #1 #2 {{\em Mon. Not. R. Astr. Soc.\/} {\bf #1}, #2} 
\def \MEM #1 #2 {{\em Mem. R. Astr. Soc.\/} {\bf #1}, #2} 
\def \PLR #1 #2 {{\em Phys. Lett. Rev.\/} {\bf #1}, #2} 
\def \PASJ #1 #2 {{\em Publ. Astron. Soc. Japan\/} {\bf #1}, #2} 
\def \PASP #1 #2 {{\em Publ. Astr. Soc. Pacific\/} {\bf #1}, #2} 
\def \NAT #1 #2 {{\em Nature\/} {\bf #1}, #2} 
\def \SAIT #1 #2 {{\em Mem.\ Soc.\ Astron.\ It.\/} {\bf #1}, #2} 
\def \MESS #1 #2 {{\em The Messenger\/} {\bf #1}, #2} 
\def \ASTRNACH #1 #2 {{\em Astron. Nach.\/} {\bf #1}, #2} 
\def \AGPSR #1 #2 {{\em ASI Special Publication\/} {\bf #1}, #2} 

\begin{opening} 
 
\title{Radioquiet isolated neutron stars: 
old and young, nearby and far away, dim and very dim} 
\author{S.B. Popov$^{1,2}$, A. Treves$^{3}$, R. Turolla$^{1}$} 
\institute{$^1$University of Padova, Department of Physics, Italy\\ 
$^2$Sternberg Astronomical Institute, Moscow, Russia\\ 
$^3$University dell'Insubria, Como, Italy} 
\date{} 
\end{opening}

\begin{document} 
 
\oddpagefooter{}{}{} 
\evenpagefooter{}{}{} 
\medskip  
 
\begin{abstract} 
 We briefly discuss the evolutionary path and observational appearance of 
isolated neutron stars (INSs) focusing on radioquiet objects. 
There are many reasons to believe that these sources are extremely 
elusive once the star surface has cooled down: their high spatial 
velocities, the long propeller stage and/or the very low accretion 
efficiency. We describe recent population synthesis models of 
close-by young INSs, highlighting the major difficulties 
encountered in the past by these simulations in reproducing the 
observed properties of known sources. As we show, 
a likely possibility is that most of the INSs in the Solar 
proximity are young (less than few Myrs)
 neutron stars born in the Gould Belt. To 
stay hot enough and sustain X-ray emission for a time $\approx 1$ 
Myr, they probably need to be low- to medium-massive, with 
$M$ less than$\sim 1.35\, M_\odot$. 
\end{abstract} 
 
\medskip 
 
\section{Introduction} 
 
Neutron stars (NSs) are among the main candidates for detection 
by new $\gamma$-ray missions, including AGILE (see eg Grenier, 
Perrot 2001 and Grenier these proceedings). Researchers usually 
resort to population synthesis modeling to derive estimates for 
the number of sources observable at present in a given energy 
band, and $\gamma$-rays are no exception (eg Gonthier et al. 2002 
and Gonthier this volume). Up to now most such studies were based 
(directly or indirectly) on the assumption that radio pulsars are 
representative of the entire Galactic NS population. 
During the last decade, however, growing evidence gathered in 
favor of the existence of radiosilent NSs, i.e. neutron stars 
which are not active radio emitters. Today the possibility that a 
significant fraction of NSs never pass through the stage of radio 
pulsars is regarded as highly plausible, as already stressed 
several years ago by Caraveo et al. (1996). This implies that a 
complete picture of NSs evolution can be obtained only by taking 
into account also radioquiet objects. The existence of a NS 
population with properties quite distinct from those of ordinary 
radio pulsars is of importance in connection with $\gamma$-ray 
sources studies, since radioquiet NSs can be bright at high 
energies, or can have evolutionary links with other $\gamma$-ray 
sources (like AXPs or SGRs). 
 
In the following we will address two main issues, the first 
concerning the paucity of detected X-ray emission from INSs, the 
second regarding the nature of a particular subclass of close-by 
INSs. The best known types of INSs like radio pulsars and active 
magnetars (AXPs and SGRs) are left out of the present discussion. 
In particular we will try to answer the two questions: 
 
\begin{itemize} 
 
\item why are INSs during all their evolution so dim? 
Especially, why there are yet no discovered old INSs powered by 
the accretion of the interstellar gas? 
 
\item What is the origin of the ``Magnificent seven'', seven dim X-ray 
sources which are associated with close-by INSs? 
 
\end{itemize} 
 
It is useful to recall at this point the four basic stages which 
characterize the interaction of an INSs with the ambient medium. 
They play a crucial role in fixing the observational properties 
of INSs and are discussed in some more detail below: {\it 
ejector\/}, {\it georotator\/}, {\it propeller\/} and {\it 
accretor\/} (see Lipunov 1992). In addition, as it is usually 
done, we term {\it coolers\/} those young NSs for which the 
surface temperature is high enough ($T\approx 10^5-10^7$~K) to make 
the star shine in the (soft) X-rays. We stress that while the 
four stages mentioned above mutually exclude each other, this is 
not necessarily true for the {\it cooler} phase. A cooler, for example, 
may be at the same time an {\it ejector}. 
 
During the {\it ejector} stage the strong outflowing momentum flux 
produced by the spinning magnetized NS prevents the surrounding 
material to cross the light cylinder radius, $R_l=c/\omega$. The 
torque exerted by the incoming matter makes the NS to spin down. 
Usually, it is assumed that the magneto-dipole formula describes 
the rate of period increase 
 
\begin{equation} 
 P\sim P_0+3\times 10^{-4}\, \mu_{30}\, t_{yrs}^{1/2}\, {\rm s}; 
\end{equation} 
here $\mu=\mu_{30}\times 10^{30}$~G~cm$^3$ is the magnetic moment 
of the NS and $P_0$ the initial period. The end of this stage occurs 
when the period reaches a 
critical value 
 
\begin{equation} 
P_E\sim 10 \, n^{-1/4}\, \mu_{30}^{1/2}\, v_6^{1/2}\, {\rm s} 
\end{equation} 
where $v=v_6\times 10^6$~cm~s$^{-1}$ is the NS velocity and $n$ 
the interstellar medium (ISM) number density. The critical period 
corresponds to the gravitational capture radius $R_G$ being equal to 
the stopping radius $R_{Sh}$, where
 
\begin{eqnarray} 
\label{rstop}
R_G & =& (2GM)/v^2\cr 
R_{Sh}&=&\left[ \frac{2\mu^2 (GM)^2 \omega^4}{\dot M v^5 c^4}\right]^{1/2}\,. 
\end{eqnarray} 
In the previous expressions $M$ is the star mass and $\dot M$ is 
the accretion rate.  
The stopping radius is determined by the equality of the 
ram pressure of the surrounding plasma ($\sim \rho v^2/2$) to the 
internal pressure of the relativistic wind 
[$\sim (\mu^2 \omega^4)/(4\pi r^2 c^4)$]. This expression is valid for 
$R_l<R_G$, which is true for all INSs which can leave the {\it 
ejector} stage (see details in Lipunov 1992). In deriving the second of
eqs. (\ref{rstop}) we assumed that the accretion rate is given by the
Bondi formula
\begin{equation}
\label{mdot}
\dot M = \frac{2\pi(GM)^2m_pn}{v^{3}} 
\sim 10^{11}\frac{n}{{v_6}^3}\, {\rm gs}^{-1}\, .
\end{equation}
We would like to stress 
that the physics governing both the spin-down rate and the critical 
period is not yet completely understood, and this has substantial 
influence on the outcome 
of population synthesis calculations. 
 
As we mentioned earlier, {\it coolers} do not represent a separate 
evolutionary stage, although they have quite distinct 
observational properties. Normally they are 
{\it ejectors} still hot enough to be visible in soft X-rays as dim 
sources. Young ($< 1$ Myr) NSs can have surface temperatures 
$T\approx 10^5-10^6$~K and luminosities $L\approx 10^{30}-10^{32}$ 
erg~s$^{-1}$ (for a review see Yakovlev et al. 1999). 
Such sources can be observed if they are relatively close ($< 1$ kpc) 
or if they are located in particular sites, typically young supernova 
remnants (SNRs), which can be targeted with deep pointings. 
 
When the {\it propeller} stage is reached matter can penetrate inside the 
light cylinder, but the fast rotating magnetosphere does not allow 
fall-down onto the NS surface and the infalling gas is 
centrifugally driven away from the star. 
As the NS spins down the centrifugal barrier disappears when the period 
becomes less than 
 
\begin{equation} 
P_A\sim 300 \, \mu_{30}^{6/7} v_6^{9/7} n^{-3/7} \, {\rm s}. 
\end{equation} 
 
After $P_A$ is crossed, the material can reach the star surface 
and the NS finally becomes an {\it accretor}. The release of potential 
energy heats the gas and produces mainly X-rays.
(However, matter can 
accumulate around the magnetosphere if the plasma cooling rate is slow in 
comparison with heating. This intermediate stage is called {\it subsonic 
propeller}  (Davies, Pringle 1981).)
  
The {\it georotator} stage in some sense is similar to the {\it accretor} 
one, although the NS spatial velocity is so high that matter is not 
gravitationally bound 
at the magnetospheric boundary. This happens whenever the star 
velocity exceeds the critical value 
 
\begin{equation} 
v_{geo}\approx 470 \, \mu_{30}^{-1/5} n^{1/10} \, {\rm km}\, {\rm s}^{-1}. 
\end{equation} 
 
 
\section{Detectability of INSs in different stages} 
 
\subsection{Coolers} 
 
\begin{figure} 
\centerline{\psfig{figure=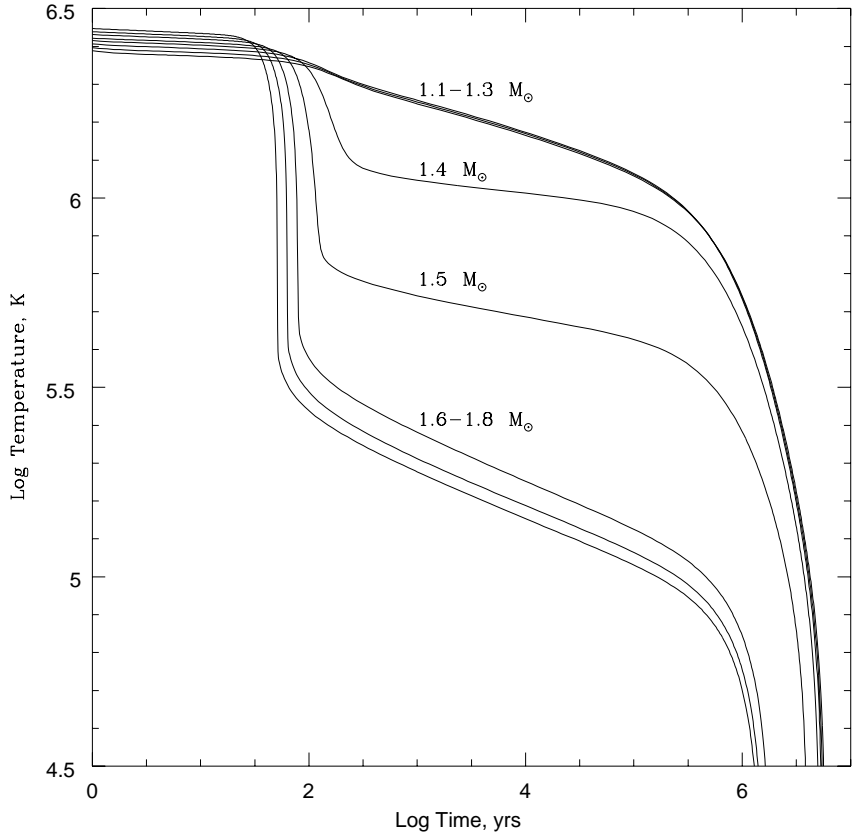,width=0.9\hsize}} 
\caption[h]{Cooling history of NSs of different masses. The redshifted 
(i.e. observed at infinity) temperature is shown 
(for a detailed discussion see Yakovlev et al. 1999; Kaminker et al. 2002)} 
\label{cool} 
\end{figure}

According to recent theoretical investigations of NS thermal 
evolution, young INSs are expected to stay hot for a relatively 
long time ($\approx 1$ Myr) if they are low-mass 
($M<1.4M_\odot$). More massive NS cool faster. 
We used cooling curves by Yakovlev et al. (1999; see 
fig.~\ref{cool}) to perform a population synthesis calculation and derive 
the Log~N~--~Log~S distribution of close-by {\it coolers}. The results 
were then compared  with observations. An important new twist of our model 
is the inclusion of INSs born in the Gould Belt in addition to those 
originating in the Galactic disk. The Gould Belt (see eg Grenier, Perrot 
2001) is a collection of stellar associations rich in young, massive 
stars. The Sun itself is embedded in the Belt which extends in a 
disk-like structure for $\sim 1$ kpc, centered at about 100 pc from 
the Sun. Due to the presence of the Belt, the rate of SNs in the Solar 
vicinity (a region a few hundred parsecs wide) during the past 
$\sim 10^7$~yrs has been higher than in an average location in the 
Galactic disk at the same distance from the Galactic center. 
Since cooling curves are strongly mass-dependent (see again 
fig.~\ref{cool}), different mass spectra for NSs has been tested.
The number of NSs was evaluated from available observations of
SN progenitors in the solar proximity and accounting for recent 
calculations of SN rates.
 
\begin{figure} 
\centerline{\psfig{figure=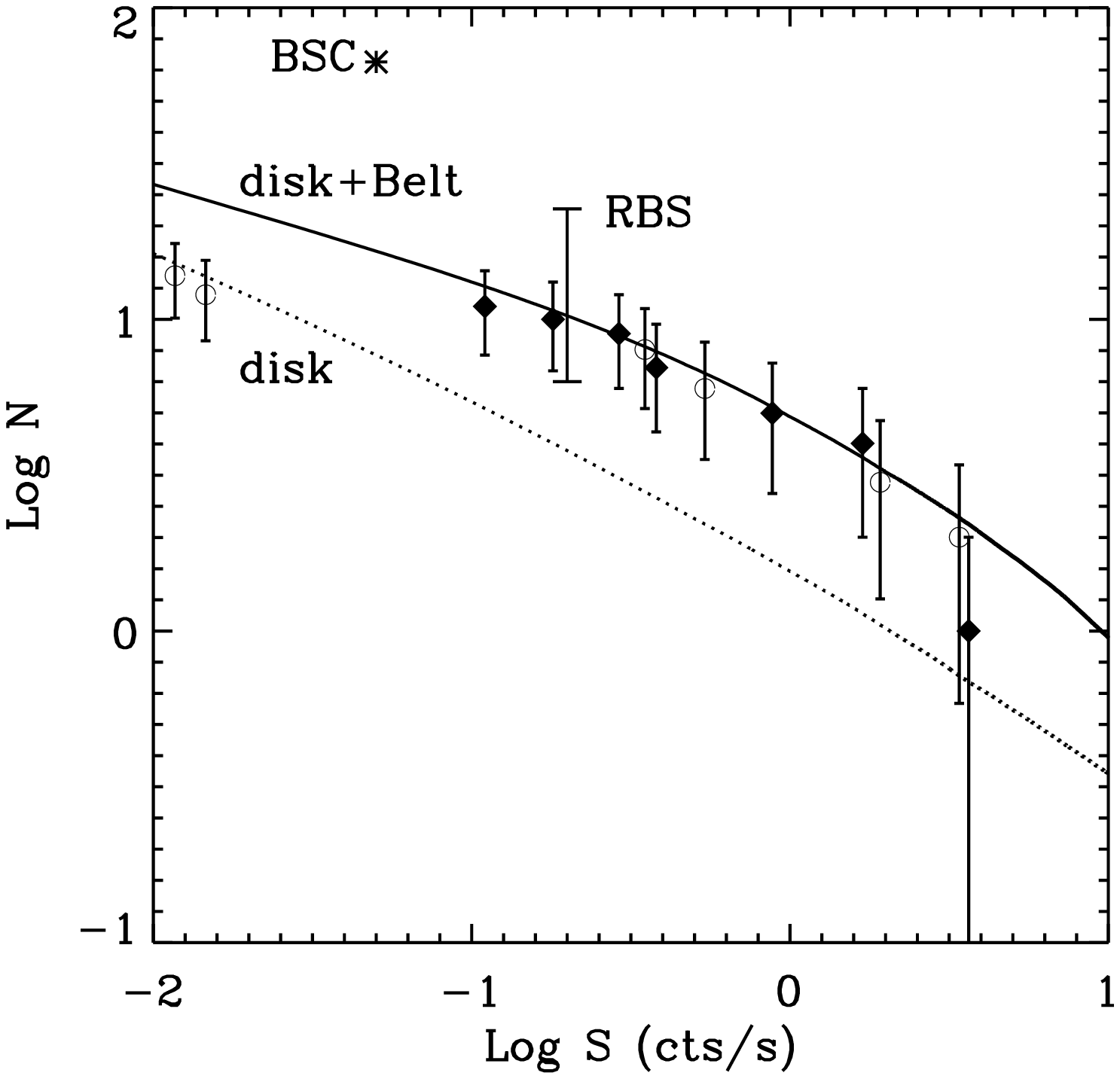,width=\hsize}} 
\caption[h]{Log~N~--~Log~S distribution for {\it coolers} (from Popov et al. 
2003b). Filled diamonds are used whenever the dimmest source at a given 
flux is one of the seven RINSs. An open circle is used for all the other 
cases. The two curves give the results of our calculation: a)
only stars from the Galactic disk contribute to the Log~N~--~Log~S 
distribution (dotted line), and b) also the contribution of the Gould Belt is 
accounted for (solid line). The labels RBS and BSC mark 
two observational limits, obtained from ROSAT data 
(RBS: Schwope et al. 1999; BSC: Rutledge et al. 2003).} 
\label{lognlogs} 
\end{figure}

The main results of our calculations are summarized in 
fig.~\ref{lognlogs}. As it can be seen comparing the two 
theoretical curves, the contribution of NSs born in the Gould 
Belt is essential in reproducing the observed Log~N~--~Log~S of 
INSs. 
 
The sample of thermally emitting INSs our model was confronted with 
(see table 1) comprises both active pulsars and radioquiet NSs. In 
particular it contains the seven soft, thermal X-ray radiosilent 
sources discovered in ROSAT pointings and known as ROSAT INSs or 
the ``Magnificent Seven''. These sources have remarkably similar 
properties that make them peculiar with respect to other X-ray bright 
INSs, like PSRs. What is special about the "Magnificent seven"? 
The most  likely explanation for these objects is that they 
are young cooling INSs, as supported by our population synthesis 
model (Popov et al. 2000b, 2003). For some reason all of them are 
radioquiet, meaning that no radio emission has yet been detected, 
although the possibility that the radio beams miss the Earth 
can not been excluded. Four objects have a measured spin period 
in the 10-20~s range quite larger than that of PSRs (see table 1). 
Moreover, none of them exhibits the  hard (non-thermal) tail which 
is typically seen in other INSs. Does this mean that the seven 
represent a distinct subpopulation of NSs? 
 
At present no definite answer can be given to this question. 
We just note that the seven ROSAT INSs are likely to be 
relatively low-massive for their temperature to be $\approx 10^6$~K 
after a relatively long time ($\approx 1$~Myr, their putative age). 
According to the cooling model we used this implies 
$M<1.35\, M_{\odot}$, so ``The Magnificent seven'' should be 
representative of the lighter part of the NS mass spectrum. 
We stress that while mass variations in the range 1-2~$M_{\odot}$ 
have little influence on other NS parameters (eg spin, accretion rate) 
they are very important for the cooling history and the internal 
structure. Correlations between the star mass and the initial 
magnetic field and/or spin period may arise soon after the proto neutron 
star has formed, eg in the fallback process. This can translate 
into a correlation between the temperature (the cooling history 
is determined by mass) and magnetic field and spin period later in the 
star evolution. On the other hand, a comparable number of NSs should be 
born with a mass above or below the critical value for fast 
cooling, $M<1.35\, M_{\odot}$, so the ``Magnificent seven'' 
are not expected to be unique objects. 
 
\begin{table} 
\caption{Local ($D<1$ kpc) population of young (age $<4.25$ Myrs) 
isolated neutron stars\vspace*{1pt}} 
{\footnotesize 
\begin{tabular}{|l||c|c|c|c|c|c|} 
\hline 
\hline 
 
{} & {} & {} &{} &{} &{} &{}\\[-1.5ex] 
Source name & Period & CR$^a$ & $\dot P$ & D & Age$^b$ & Refs 
\\[1ex] 
            &   s     & cts/s & $10^{-15}$ s/s& kpc   & Myrs     & 
\\[1ex] 
 & & & & & & \\[1ex] 
\hline 
RINSs & & & & & & \\[1ex] 
RX J1856.5-3754            &  ---  & 3.64  &  ---&0.117$^e$&$\sim0.5$& 
[1,2]\\[1ex] 
RX J0720.4-3125                 & 8.37 & 1.69  &$\sim 30-60$& ---&---& 
[1,3]\\[1ex] 
RX J1308.6+2127 & 10.3 & 0.29  & --- & --- & --- & 
[1.4] \\[1ex] 
RX J1605.3+3249       &  ---  & 0.88  & --- & --- & --- & 
[1]\\[1ex] 
RX J0806.4-4123                 &  11.37  & 0.38  & --- & --- & --- & 
[1,5]\\[1ex] 
RX J0420.0-5022                 &  22.7   & 0.11  & --- & ---   & --- & 
[1]\\[1ex] 
RX J2143.7+0654 &  ---    & 0.18  & --- & ---  & --- & 
[6]\\[1ex] 
 & & & & & & \\[1ex] 
\hline 
 Geminga type& & & & & & \\[1ex] 
PSR B0633+17            & 0.237 & 0.54$^d$ &10.97&0.16$^e$&0.34& 
[7]\\[1ex] 
3EG J1835+5918 & ---   & 0.015    & --- & ---  &  --- & 
[8]\\[1ex] 
 & & & & & & \\[1ex] 
\hline 
 Thermally emitting PSRs & & & & & & \\[1ex] 
PSR B0833-45    & 0.089 & 3.4$^d$  & 124.88 & 0.294$^e$ & 
0.01& 
[7,9,10]\\[1ex] 
PSR B0656+14          & 0.385 & 1.92$^d$ &  55.01 & 0.762$^f$ & 0.11 
& 
[7,10]\\ [1ex] 
PSR B1055-52          & 0.197 & 0.35$^d$ &   5.83 & $\sim 1^c$ & 
0.54& 
[7,10]\\ [1ex] 
PSR B1929+10          & 0.227 & 0.012$^d$& 1.16 &  0.33$^e$  & 3.1& 
[7,10]\\[1ex] 
 & & & & & & \\[1ex] 
\hline 
Other PSRs & & & & & & \\[1ex] 
PSR J0056+4756  & 0.472 & --- & 3.57 &  0.998$^f$  & 2.1& 
[10]\\[1ex] 
PSR J0454+5543  & 0.341 & --- & 2.37 &  0.793$^f$  & 2.3& 
[10]\\[1ex] 
PSR J1918+1541  & 0.371 & --- & 2.54 &  0.684$^f$  & 2.3& 
[10]\\[1ex] 
PSR J2048-1616  & 1.962 & --- & 10.96&  0.639$^f$  & 2.8& 
[10]\\[1ex] 
PSR J1848-1952  & 4.308 & --- & 23.31&  0.956$^f$  & 2.9& 
[10]\\[1ex] 
PSR J0837+0610  & 1.274 & --- & 6.8  &  0.722$^f$  & 3.0& 
[10]\\[1ex] 
PSR J1908+0734  & 0.212 & --- & 0.82 &  0.584$^f$  & 4.1& 
[10]\\[1ex] 
\hline 
\hline 
\multicolumn{7}{l}{ 
$^a$) ROSAT PSPC count rate}\\ 
\multicolumn{7}{l}{ 
$^b$) Ages for pulsars are estimated as $P/(2\dot P)$,}\\ 
\multicolumn{7}{l}{ 
      for RX J1856 the estimate of its age comes from kinematical 
      considerations.}\\ 
\multicolumn{7}{l}{ 
$^c$) Distance to PSR B1055-52 is uncertain ($\sim$ 0.9-1.5 kpc)}\\ 
\multicolumn{7}{l}{ 
$^d$) Total count rate (blackbody + non-thermal)}\\ 
\multicolumn{7}{l}{ 
$^e$) Distances determined through parallactic measurements}\\ 
\multicolumn{7}{l}{ 
$^f$) Distances determined with dispersion measure}\\ 
\multicolumn{7}{l}{ 
[1] Treves et al. (2000) ; [2] Kaplan et al. (2002); [3] Zane et al. (2002);}\\ 
\multicolumn{7}{l}{ 
[4] Hambaryan et al. (2001); [5] Haberl, Zavlin (2002); [6] Zampieri et al. 
(2001);}\\ 
\multicolumn{7}{l}{ 
[7] Becker, Tr\"umper (1997); [8] Mirabal, Halpern (2001); [9] Pavlov et al. 
2001;}\\ 
\multicolumn{7}{l}{ 
[10] ATNF Pulsar Catalogue (see Hobbs et al. 2003)}\\ 
\hline 
\end{tabular} } 
\vspace*{-13pt} 
\end{table} 
 
To conclude this subsection: 
the "Magnificent seven" can be young {\it coolers} from the Gould Belt. 
We predict that not more than several tens of young INSs 
in future can be identified in 
ROSAT data at low galactic latitudes and some of them can also be among EGRET 
unidentified sources. There is a possibility, that due to processes 
accompaning NS birth there is a correlation between mass of compact object 
and its initial magnetic and spin period.

\subsection{Ejectors and Georotators} 
 
According to the simulations by Popov et al (2000a)  
the large majority of INSs (about 90 \%) never leave the
{\it ejectors} stage if the star magnetic field is $\sim 10^{12}$~G 
(or lower) and does not decay. 
Those calculations referred to a NS 
population with average velocity $\sim 200-300$~km~s$^{-1}$; 
since $P_E$ scales as $v^{1/2}$ (see eq. [\ref{mdot}]),
for higher mean velocities the fraction of {\it ejectors} is even higher. 
We draw the reader's attention to the fact that the number of observable 
accreting INS is very low for high star velocities not because the 
accretion rate (and hence the luminosity)
is low, but simply because INSs cannot reach the stage 
of accretion. There has been some confusion in literature about this point. 
Many investigators attributed the lack of observed {\it accretors} in
ROSAT data, in contradiction with the initial predictions by 
Treves, Colpi (1991) and others, to the low accretion luminosity 
of fast-moving NSs. This would imply that the Galaxy 
is filled with very dim {\it accretors}, which is not the case ! 
Even leaving aside further effects that can influence the duration and/or
the efficiency of accretion (eg a long {\it subsonic propeller} stage, 
a below-Bondi rate; see discussion in section \ref{accretors}) 
the number of {\it accretors} (of all luminosities !) 
goes below a few percent if the average velocity is higher than $\approx
200$~km~s$^{-1}$. In this case, the number of accreting INSs is about
two orders of magnitude below the original prediction by Treves, Colpi (1991). 

Let us very briefly touch some difficulties in describing {\it ejectors} 
in population synthesis models. 
Most  population synthesis studies 
of INSs are based on some simple model assumptions which can be a significant 
oversimplification. 
For example, transition from 
{\it ejector} stage to {\it propeller} or {\it georotator} stage 
is assumed to appear when surrounding plasma can penetrate inside 
the light cylinder (see Lipunov 1992). However the reality can be
more complicated (see for example discussion in Michel 2003), and
not $R_l$ can be the most important scale. 
 Also  we actually do not understand well enough the mechanism 
of energy losses on the {\it ejector} stage, 
we do not know how the braking index $n$ evolves, we do not know 
if magnetic and spin axis align or contr-align (see for example Beskin et al. 
1993 and Regimbau, de Freitas Pacheco 2001), we do not know if the magnetic 
field significicantly decay during the {\it ejector} stage and how it is 
connected with spin evolution (see Konenkov, Geppert 2001). 
 To summarize: we do not know how do {\it ejectors} live and die... 
So, the building of the population synthesis of INSs, which starts with the 
{\it ejector} stage, has not very solid basis! 
 
We stress once more that the {\it ejector} phase does not necessarily
coincide with a NS being an active radio pulsar. {\it Ejectors} comprise,
in fact, also a) pulsars that already crossed the death valley
and, b), NSs which never become radio loud. This means that the vast
majority of Galactic NSs is at present inaccessible to observations.  
There has been several proposals aiming at the detection of
these high-velocity INSs. Rutledge (2001) developed the idea by 
Harding, Leventhal (1992) that INSs in the {\it georotator} stage 
during which full-fledged accretion can not occur because the magnetospheric 
radius is larger than the gravitational capture radius) can still 
accrete warm matter from the ISM. However, the scenario 
proposed by Rutledge requires ultra-high ($>10^{15}$~G) magnetic fields 
during the entire lifetime of a NS which sounds a rather doubtful
assumption. Field decay will make 
this stage (called MAGAC -- magnetically accreting) very short. Toropina 
et al. (2001) and Romanova et al. (2001) proposed 
that activity in a long magnetospheric ``tail''  can lead to observable 
consequences due to reconnection of the magnetic field lines, as 
observed in numerical experiments. 

We note that the astrophysics of high-velocity magnetized INSs is far from 
being completely understood. Up to very recent times,
the detection of such dim, high-energy sources was simply bejond imagination.
The role of the star velocity in shaping the magnetosphere is definitely
a very important issue in establishing their observational properites.
``Mixed'' stages, in which the relevant lenghtscales (eg the Alfv\`en radius) 
up-stream and down-stream  are different because of the highly non-symmetric 
interaction with the ISM, as discussed in Toropina et al. (2001) and 
Romanova et al. (2001) deserves a more thorough investigation. 
 
\subsection{Propellers} 
\label{propel} 
As discussed in the previous section, the number of {\it propellers} is 
very low just because virtually all NSs are {\it ejectors} for a 
non-decaying magnetic field. If the field decays, however, the situation 
can be the opposite (eg Colpi et al. 1998; Livio et al. 1998; Popov et al. 
2000a; Popov, Prokhorov 2000). 
INSs can enter the region of the parameter space where the magnetic 
field is low enough to make the spin evolution very slow, and, at the same 
time, the period is not too long to allow accretion. For example, 
if field decay freezes at about $10^{8}-10^{10}$~G, 
and the decay time-scale is $\sim 10^8$~ yrs, a NS 
with initial field $\sim 10^{12}$~G never reaches a stage where 
$R_A=R_{co}$ (see Popov, Prokhorov 2000). 
 
The star enters the {\it propeller} stage with $P=P_E$ and starts to 
spin down intensively. In a relatively short time ($\approx 10^6$~yrs for 
$B\sim 10^{12}$~G) the spin period increases up to $P=P_A$. At that moment $
R_A=R_{co}$ and if matter can penetrate the magnetosphere, it can fall down 
onto the star surface. However, even if $R_A<R_{co}$ a NS can avoid the 
accretion stage if cooling in the atmosphere above $R_A$ is not effective. 
Under such conditions,  in fact, no instabilities can set 
in the hot plasma preventing matter to enter the magnetosphere (Davies, 
Pringle 1981). It is the so called {\it subsonic propeller} stage. Recently 
Ikhsanov (2003) re-investigated this issue in connection with INSs. 
His main conclusion is that in a realistic situation an INS 
spends a significant part  of its life ($ > 10^9$~yrs) 
in this stage during which only a small amount of matter can diffuse inwards
and reach the star surface. If this situation is realized in nature than even 
low-velocity INSs may never become bright {\it accretors}. 
 
\begin{figure} 
\centerline{\psfig{figure=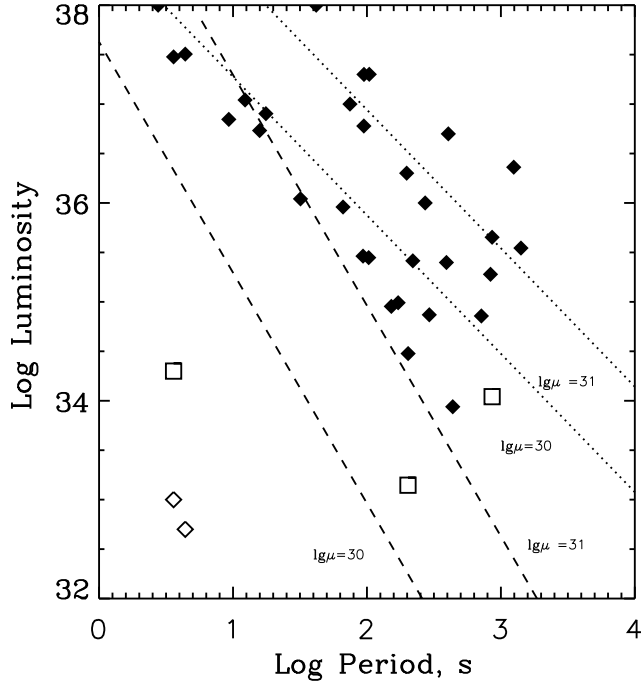,width=10cm}} 
\caption[h]{ 
Period vs. luminosity for Be/X-ray binaries; data taken mainly 
from Haberl, Sasaki (2000). Open symbols correspond to the quiescent state 
of the X-ray pulsar. Squares represent three sources in quiescence from
which pulsations were observed (4U 0115+63 -- Campana et al. 2001; 
RX J0440.9+4431 and RX J1037.5-564 -- Reig, Roche 1999). 
Open diamonds show objects without pulsations, which are supposed to be 
in the {\it propeller} state (4U 0115+63 and V0332+53 -- Campana et al. 2002). 
The two dashed lines correspond to the critical period, 
$P_A=2^{5/14}\pi (GM)^{-5/7} (\mu^2/\dot M)^{3/7}$, for two values 
of the magnetic moment, $\mu=10^{30}$~G~cm$^3$ 
and $10^{31}$~G~cm$^3$. The 
two dotted lines correspond to subsonic propeller~--~accretor 
transition for the same two values of the magnetic moment which
occurs at 
$P_{crit}=81.5 \mu_{30}^{16/21} L_{36}^{-5/7}$ according to Ikhsanov (2003). 
We note that the multiplicative coefficient in Ikhsanov's formula is 
larger than in the classical formula of 
Davies, Pringle (1981) by a factor $\sim7.5$.\\
For most objects luminosities correspond to outburst phases, as is apparent
from the plot, where the points are shifted from the equilibrium line.
Equilibrium should correspond to some "average" over the 
orbital period luminosty.} 
\label{psr} 
\end{figure}

One can try to investigate the existence of {\it subsonic propeller} using 
X-ray binaries (probably, Be/X-ray systems are the closest analogue to 
accreting INSs). However, in X-ray binaries the plasma density is higher 
than in INSs and results cannot be applied directly. 
In fig.~\ref{psr} we show data for X-ray pulsars in binary systems 
with known spin periods and luminosities. We used also long period pulsars in 
Be systems, so they are expected to be far from spin equilibrium
(see Lipunov 1992 for details). 
As it can be seen, data are in better agreement with the ``normal'' critical 
period, than with the subsonic one (the reason could be the higher plasma 
density in these systems). Cases when one can suspect X-ray systems to be 
at the {\it propeller} stage (filled diamonds in the figure) lie much below 
the others. 
 
If the {\it subsonic propeller} stage is taking place then accreting INSs 
should have very long spin periods, probably absence of periodic pulsations 
can be a signature of an {\it accretor} (on spin period of INSs at the stage 
of accretion see Prokhorov et al. 2002). 
We note, that an INS can be an accreting object with period 
about 10 seconds only if its field significantly decayed (Konenkov, Popov 
1997; Wang 1997). However, at least for one of the seven ROSAT INSs spin-down
implying a field $B\sim 2-6\times 10^{13}$~G was measured. This provides
further support, in addtion to the large measured spatial velocities 
for same of them, to reject 
the idea that the seven are powered by accretion 
(see a review of recent observational data in Haberl 2003). 
 
\subsection{Accretors} 
\label{accretors} 
The first ideas about INSs accreting from the ISM appeared in the early 70's 
(eg. Ostriker et al. 1970; Shvartsman 1971). At that time, however,
it  went nearly unnoticed since no detector was available
to verify this idea. It was only 20 years later that, thanks to ROSAT 
sensitivity, that it became possible to detect such dim, soft sources.
This prompted new interest and Treves and Colpi (1991) were the first
to estimate the number of accreting INSs detectable with ROSAT.
Their calculations predicted that a large number of accreting INSs should
be visible in ROSAT exposures, basing on three main assumptions 
\begin{itemize} 
\item a large fraction of low velocity objects
\item Bondi accretion rate
\item most of NSs spend a large part of their lives as {\it accretors}. 
\end{itemize}

Up to now {\it no} strong candidates for this class of sources have been 
found. The reasons for explaining the paucity of accreting 
INSs boils down to essentially to two: a) there are almost no 
{\it accretors}, and b) the accretion luminosity is much lower than what
implied by the Bondi formula, using conventional efficiency for conversion of 
gravitational potential energy into radiation ($\eta\sim GM/Rc^2$). 
As we discussed above, there are good reasons to believe that not all INSs 
become {\it accretors}, and that even those that do so spend most of 
their lives in earlier stages ({\it ejector, propeller}). 
The NS velocity distribution alone is enough to explain 
the lack of bright ($>0.1$~ROSAT~cts~s$^{-1}$) {\it accretors}. 
The possibility that the propeller stage is long (longer than  a few Myrs
as assumed, for example, in Popov et al. 2000a,b) makes the statement 
above even stronger. 

In addition, recent investigations strongly support the idea that 
the Bondi rate is just an upper limit which is rarely realized in nature. 
There are three main reasons for which $\dot M\ll \dot M_{Bondi}$:  
pre-heating, magnetic effects and a low-efficiency accretion flow.
As noted by Blaes et al. (1995) the UV/X-ray flux coming form the star
surface will heat the incoming matter producing an increase of the gas 
temperature and hence of the local sound speed $c_S$. If the star velocity is 
lower than $c_S$ the latter is governing the accretion rate and
accretion is reduced. This effect is important for slow INSs and can reach 
1.5 orders of magnitude for $v=20$~km~s$^{-1}$. Recent 2D MHD simulations
have shown that accretion onto a rotating dipole is substantially different
from that onto an unmagnetized star (Romanova et al. 2003; 
Toropina et al. 2003) Magnetic effects scales with the magnetic field
strenght. For lower field they are less pronounced. If the field decays 
significantly, these effects are not so serious, otherwise the accretion 
rate is reduced by more than one order of magnitude. 
Finally, recent 2D and 3D simulations of accretion flows show that 
convection can reduce the accretion rate by orders of magnitude (see 
Perna et al. 2003 and references therein). However, when
applied to INSs the formula 
 
\begin{equation}
\dot M \sim \left( \frac{R_{in}}{R_{out}} \right)^p \dot M_{Bondi}, \, 
p\sim 1/2, 
\end{equation}
which is obtained in the case of black hole accretion, 
does not provide a strong constraint since $R_{in}\approx R_A$ 
and $R_{out}\approx R_G$, so, typically, 
$(R_A/R_G)^{1/2}$ is only about 0.1. 
We note also that in binaries where a NS is the accreting compact object
(at accretion rates $>10^{15}$~g~s$^{-1}$) the Bondi formula seem to
work. All the data on luminosity are in good correnspondence with 
estimates of $\dot M_{Bondi}$, as derived from the  wind parameters and 
geometry of these systems. 
 
There is a hope to observe isolated {\it accretors} at specific sites. for 
example in globular clusters (GCs). 
It was suggested by Pfahl and  Rappaport (2001) that some dim X-ray 
sources detected by Chandra in GCs can be old INSs accreting ISM. 
Simple evolutionary calculations (Popov, Prokhorov 2002) support this idea. 
A cluster with mass $\sim 10^5 M_{\odot}$ can have one accreting INS. 
However, this is strongly dependent on the amount and the properties of the 
ISM in GCs which are not well constrained at present. 
Additionly, young INSs can be accreting due to so called remnant (or 
fall-back) discs (for example Chatterjee et al. 2000, Alpar 2003). We will not 
discuss this possibility further here. 
 
\begin{figure} 
\centerline{\psfig{figure=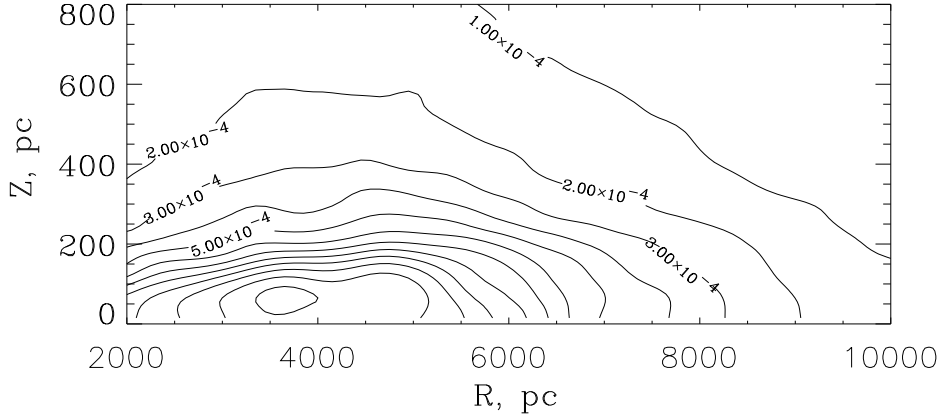,width=\hsize}} 
\caption[h]{Spatial distribution of INSs in the Galaxy (from Popov et al. 
2003b). The data was calculated by a Monte-Carlo simulation 
of $>$~10000 individual 
tracks on a fine grid (10 pc in $z$ direction and 100 pc in $R$ direction). 
Kick velocity was assumed following Arzoumanian et al. (2002). 
NSs were born in the thin disk with semithickness 75 pc. 
No NS born inside $R=2$~kpc and outside $R=16$~kpc were taken into account. 
NS formation rate was assumed to be constant in time 
and proportional to the square of the ISM 
density at the birthplace. Results were normalized to have in total 
$5\times 10^8$ NSs born in the described region. 
Density contours are shown with a step 
$0.0001$~pc$^{-3}$. At the solar distance from the center close to the 
Galactic plane the NS density is about $2.8\times 10^{-4}$~pc$^{-3}$. 
} 
\label{dens} 
\end{figure} 
 
\section{Conclusion} 
 
When Baade and Zwicky (1934) proposed the idea of neutron stars 
there were not many hopes to observe these objects. It was realized 
since the very beginning that the very small radius would make them
very dim thermal emitters, unless their temperature is extremely large.
It was only more than 30 years later that the existence of radio pulsars
was recognized and this paved the way to the discovery of a large
number of INSs. However, if an INS is radioquiet (or its radio beam misses 
the Earth), than  the original conclusion still holds: isolated NSs
are very faint objects, if they emit at all.  Young INSs (ie {\it coolers}) 
are, by far, the most numerous class of 
known radioquiet INSs (including also the central sources in SNRs). 
These objects are doomed to fade in a very short time ($\approx 10^5$~yrs) 
if their masses are larger than some critical value which is about 
$1.35\, M_{\odot}$. Anyway, they will cool below the detection limit 
in a few million years. 
Old INSs can be resurrected due to accretion. So the only 
possibility to observe an old INSs is to see it as an {\it accretor}. 
Otherwise they are extremely dim old {\it ejectors} 
with negligible spin-down losses, 
 {\it propellers} or {\it georotators} from which there are not much 
hopes for any observable activity. 
Here we summarize the main  reasons for which INSs older than 
$\approx 10^6-10^7$~yrs are expected to be substantially undetectable. 

1) The high spatial velocity: a) if the magnetic field is constant with value 
typical of normal radio pulsars, then INSs with $v>100$~km~s$^{-1}$ spend 
all their lives as {\it ejectors}; b) for a decaying field there can be 
realistic values of the parameters for which INSs freeze at the 
{\it propeller} stage (see section \ref{propel}); c) for fast decay or 
very low initial field, when the magnetospheric barrier does not exist, 
accretion rate is too low for high velocity INSs to produce an observable 
object (see section \ref{accretors}).

2) The long propeller stage: a NS 
can spend nearly all its life as a {\it 
propeller} due to magnetic field decay. Also, there is the possibility 
that a NS can spend billions of years at the so-called 
{\it subsonic propeller} stage until it spins down to very long 
periods. This stage too is expected to be a very dim one.

3) The low accretion efficiency: even if a NS comes to the stage of 
accretion there are many reasons to expect a very low luminosity. 
Among these are: a) a very low accretion rate realistic average velocity
of the NS population;  b) preheating, which decreases the accretion rate; 
c) MHD effects which prevent material to reach the star surface;
d) low accretion rate due to turbulence. 
 
Anyway, at least several radioquiet INSs are already observed 
(the "Magnificent seven", Geminga and 3EG J1835+59, 
plus compact X-ray sources in SNRs and some others), 
and their number will grow in future 
also thanks to new $\gamma$-ray missions like AGILE and GLAST. 
As can be seen from fig.~\ref{dens} there are about $(2-3)\times 10^{-4}$ 
per cubic parsec around the Sun, so there are still a lot of objects to 
discover. 
 
\acknowledgements 
SP wants to thank Mikhail Prokhorov for discussions and collaboration.

\end{document}